\documentclass[RNAAS]{aastex63}

\usepackage{fontawesome}

\newcommand{\arf}[1]{\texttt{arf}}
\newcommand{\rmf}[1]{\texttt{rmf}}

\newcommand\simon[1]{{\color{orange} #1}}

\received{June 1, 2019}
\revised{January 10, 2019}
\accepted{\today}
\submitjournal{RNAAS}

\shorttitle{}
\shortauthors{Rhea et al.}
\usepackage{amsmath}

\graphicspath{{./}{figures/}}
\usepackage{comment}
\usepackage{physics}
\begin{document}

\title{Updates to LUCI: A New Fitting Paradigm Using Mixture Density Networks}

\correspondingauthor{Carter Lee Rhea}
\email{carterrhea@astro.umontreal.ca}

\author[0000-0003-2001-1076]{Carter Rhea}
\affiliation{Département de Physique, Université de Montréal, Succ. Centre-Ville, Montréal, Québec, H3C 3J7, Canada}
\affiliation{Centre de recherche en astrophysique du Québec (CRAQ)}

\author[0000-0001-7271-7340]{Julie Hlavacek-Larrondo}
\affiliation{Département de Physique, Université de Montréal, Succ. Centre-Ville, Montréal, Québec, H3C 3J7, Canada}

\author[0000-0002-5136-6673]{Laurie Rousseau-Nepton}
\affiliation{Canada-France-Hawaii Telescope, Kamuela, HI, United States}

\author[0000-0002-1755-4582]{Simon Prunet}
\affiliation{Canada-France-Hawaii Telescope, Kamuela, HI, United States}
\affiliation{Laboratoire Lagrange, Universit\'e C\^ote d’Azur, Observatoire de la C\^ote d’Azur, CNRS, Parc Valrose, 06104 Nice Cedex 2, France}

\begin{abstract}
LUCI is an general-purpose spectral line-fitting pipeline which natively integrates machine learning algorithms to initialize fit functions \citep[see][for more details]{rhea_machine_2020,rhea_luci_2021}. LUCI currently uses point-estimates obtained from a convolutional neural network (CNN) to inform optimization algorithms; this methodology has shown great promise by reducing computation time and reducing the chance of falling into a local minimum using convex optimization methods. In this update to LUCI, we expand upon the CNN developed in \cite{rhea_machine_2020} so that it outputs Gaussian posterior distributions of the fit parameters of interest (the velocity and broadening) rather than simple point-estimates. Moreover, these posteriors are then used to inform the priors in a Bayesian inference scheme, either \texttt{emcee} or \texttt{dynesty}. The code is publicly available at \href{https://github.com/crhea93/LUCI}{\faicon{github} crhea93:LUCI}.
\end{abstract}


\section{Methodology}

\subsection{Mixture Density Networks}
Mixture density networks combine standard neural networks with mixture density models to compute a posterior distribution of the target variables $p(\Theta|X)$ conditioned upon the initial inputs, X, where $\Theta$ represents the target variables. In this manner, they provide uncertainty estimates on the target variables. Following the procedure outlined in \cite{bishop_mixture_1994}, we assume the output distribution of a single variable has the form of a mixed density model with a Gaussian kernel:
\begin{equation}
    p(x) = \sum_{j=0}^{m-1} \pi_j p_j(y,\nu) = \sum_{j=0}^{m-1} \pi_j \frac{1}{\sqrt{2\pi\sigma_j}}\exp{\frac{-(y-\mu_j)^2}{2\sigma_j^2}}
\end{equation}

Moreover, we assume that a single Gaussian distribution is sufficient to describe the distributions of velocities and broadening parameters. Hence, we assume a posterior distribution of each parameter of the following form:

\begin{equation}
    p_\nu(y) = \frac{1}{\sqrt{2\pi\sigma_j}}\exp{\frac{-(y-\mu_j)^2}{2\sigma_j^2}}
\end{equation}

Therefore, our network must output a total of four variables: $\mu_v$, $\mu_{\Delta v}$, $\sigma_v$, and $\sigma_{\Delta v}$ where $\mu_v$ is the mean of the velocity distribution, $\sigma_v$ is the Gaussian standard deviation of the velocity distribution, $\mu_{\Delta v}$ is the mean of the velocity dispersion distribution, and $\sigma_{\Delta v}$ is the Gaussian standard deviation of the velocity dispersion distribution. 
The network is trained using standard backpropagation techniques, so a proper loss function must be chosen; we use the negative log-likelihood function, which makes the problem equivalent to a maximum likelihood estimate \citep{bishop_mixture_1994}.

We adapt the existing CNN described in \citep{rhea_machine_2020} to be a single-phase mixture density model (i.e. assuming a single Gaussian distribution is sufficient to describe the output variables). Our \texttt{tensorflow} (\citealt{abadi_tensorflow_2015}) implementation can be found at \url{https://github.com/sitelle-signals/Pamplemousse}. In \texttt{LUCI}, the mean values, $\mu_\Theta$, are used as the best estimates and the optimization problem \footnote{Meaning in the mode in which \texttt{LUCI} is using a convex optimization routine for Maximum \textit{A Posteriori} and not a Bayesian sampler.} is constrained using bounds taken from the 3-$\sigma$ values (i.e. 3 times the $\sigma_\Theta$ calculated by the network).

\subsection{Bayesian Inference}

In addition to standard fitting, \texttt{LUCI} has a Bayesian inference implementation to determine uncertainties on the fit parameters. Previous to this update, the priors were non-informative and covered a reasonable range of values; however, this led to longer calculation time and less constrained uncertainties. Therefore, in this update, we have modified \texttt{LUCI} to use the posterior distributions found by the MDN as prior distributions. Hence, the priors on the velocity and broadening are described by the parameters obtained by the MDN. However, we note that this methodology could lead to potential biases since the same data is used to calculate the prior and posterior despite the differing methodologies. We, therefore,  propose two modifications to reduce the potential biases: use the mean and sigmas calculated by the MDN to create an informed uniform prior centered at the mean with a span on either side of 3-$\sigma$ or keep uninformed priors while using the MDN-calculated distribution as the MCMC's transition distribution. The current version of \texttt{LUCI} includes the first of these two options.
An MCMC Bayesian inference pipeline, \texttt{emcee}, has been implemented (\citealt{foreman-mackey_emcee_2013}) in addition to a nested sampling pipeline, \texttt{dynesty} (\citealt{skilling_nested_2004}; \citealt{skilling_nested_2006}; \citealt{speagle_dynesty_2020}).

\section{Results \& Discussion}
\begin{figure}
    \centering
    \includegraphics[width=0.9\textwidth]{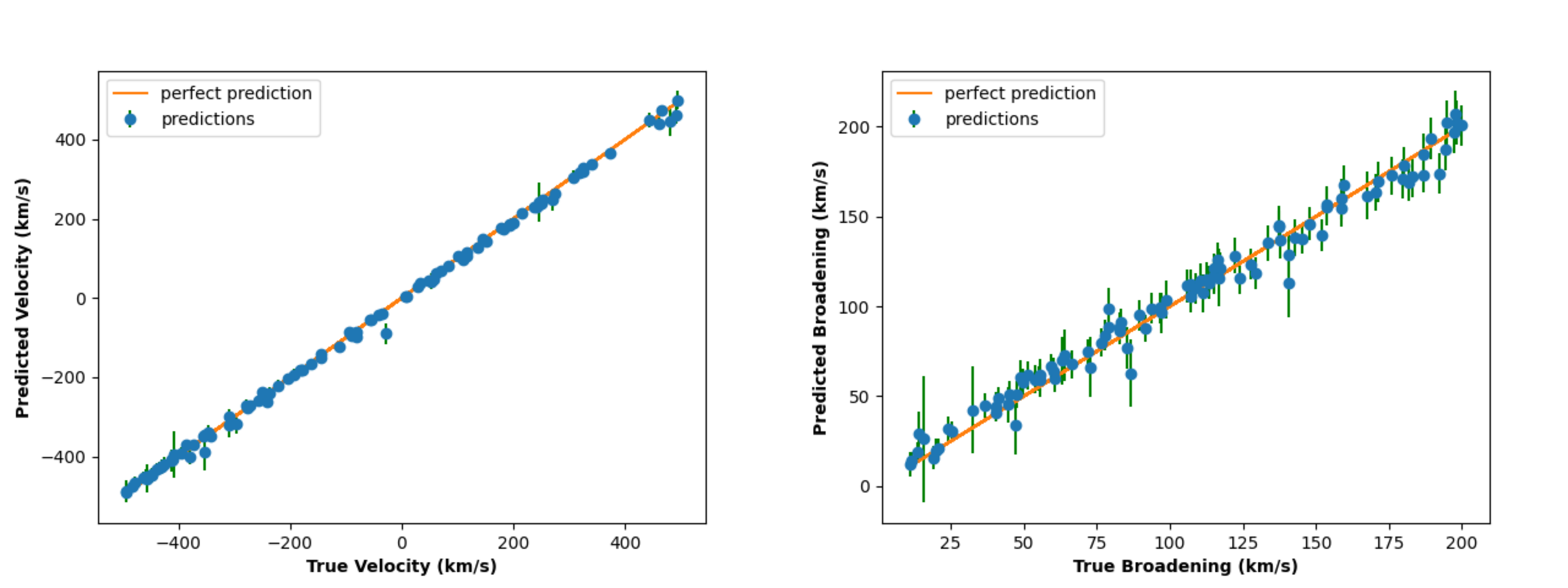}
    \caption{Outputs of a sample of 100 points randomly selected from the test set. The velocity (left) and broadening (right) results are shown. The mean of the output posterior for each spectrum is represented by a blue dot while the 1-$\sigma$ errors are shown as green error bars. The orange line represents perfect estimates (i.e. the network correctly predicts the true value).}
    \label{fig:mdn}
\end{figure}

Figure \ref{fig:mdn} shows the predicted values and associated 1-$\sigma$ errors for 100 randomly chosen test spectra at R$\approx$5000 for both the velocity (in km/s) and the broadening (in km/s), respectively. The plots reveal that the network predicts distributions centered on the actual value with minor (on the order or 10-20 km/s in both cases) uncertainties.  Using the MDN reduces uncertainties by up to 400\% when compared to the standard CNN implementation.

Adapting the original convolutional neural network to output posterior distributions rather than single-point estimates represents a new paradigm in standard fitting procedures. Once the distributions are calculated, they can be used either as bounds for an optimization-based fit (such as least squares) or as priors in Bayesian inference. This novel paradigm has been implemented in \texttt{LUCI}; users have the option of switching between the standard and updated methodology by setting the \texttt{mdn} boolean to either \texttt{False} or \texttt{True}, respectively.

\texttt{LUCI} is distributed freely and can be downloaded from our GitHub page: \url{https://github.com/crhea93/LUCI}\faicon{github}. This page includes installation instructions for Linux, Mac, and Windows operating systems in addition to a contributors guide, citation suggestions, and example Jupyter notebooks. Our online documentation (\url{https://crhea93.github.io/LUCI/index.html}) includes pages on the mathematics behind \texttt{LUCI}, descriptions of uncertainty calculations, extensive examples, API documentation, and a FAQ/errors section.

\acknowledgments

The authors would like to thank the Canada-France-Hawaii Telescope (CFHT), which is operated by the National Research Council (NRC) of Canada, the Institut National des Sciences de l'Univers of the Centre National de la Recherche Scientifique (CNRS) of France, and the University of Hawaii. The observations at the CFHT were performed with care and respect from the summit of Maunakea, which is a significant cultural and historic site.

C. L. R. acknowledges financial support from the physics department of the Universit\'{e} de Montr\'{e}al, IVADO, and le fonds de recherche -- Nature et Technologie.

J. H.-L. acknowledges support from NSERC via the Discovery grant program, as well as the Canada Research Chair program.

\software{astropy \citep{collaboration_astropy_2013}, scipy \citep{virtanen_scipy_2020}, numpy \citep{harris_array_2020}, matplotlib \citep{hunter_matplotlib_2007}, ipython \citep{perez_ipython_2007}, tensorflow \citep{abadi_tensorflow_2015}, keras \citep{chollet_keras_2015}, emcee \citep{foreman-mackey_emcee_2013}}

\bibliography{Luci}{}
\bibliographystyle{aasjournal}

\end{document}